\documentclass[a4paper,12pt]{article}
\usepackage{jcappub}
\usepackage[symbol]{footmisc}
\usepackage{amsmath,amssymb}
\usepackage{revsymb}
\usepackage{orcidlink}

\newcommand{\nc}{\newcommand*} 
\nc{\red}[1]{\textcolor{red}{#1}}
\nc{\Eq}[1]{Eq.~\eqref{#1}}     
\nc{\Fig}[1]{Fig.~\ref{#1}}     
\nc{\Table}[1]{Table~\ref{#1}}  
\nc{\Sec}[1]{Sec.~\ref{#1}}     
\def\({\left(}
\def\){\right)}
\def\[{\left[}
\def\]{\right]}
\def\e{\begin{equation}}
\def\q{\end{equation}}
\def\m{\begin{eqnarray}}
\def\n{\end{eqnarray}}

\begin{document}

\title{Constraints on {Inflation with} Null Energy Condition Violation from Advanced LIGO and Advanced Virgo's First Three Observing Runs}

\author[a,b]{Zu-Cheng~Chen\orcidlink{0000-0001-7016-9934},}
\author[c,d,*]{Lang~Liu\orcidlink{0000-0002-0297-9633},\note{Corresponding author.}}

\affiliation[a]{Department of Physics and Synergetic Innovation Center for Quantum Effects and Applications, Hunan Normal University, Changsha, Hunan 410081, China}
\affiliation[b]{Institute of Interdisciplinary Studies, Hunan Normal University, Changsha, Hunan 410081, China}
\affiliation[c]{Department of Astronomy, Beijing Normal University, Beijing 100875, China}
\affiliation[d]{Advanced Institute of Natural Sciences, Beijing Normal University, Zhuhai 519087, China}

\emailAdd{zuchengchen@hunnu.edu.cn}
\emailAdd{liulang@bnu.edu.cn}

\abstract{
The null energy condition (NEC) is a cornerstone of general relativity, and its violation could leave observable imprints in the cosmic gravitational wave spectrum. Theoretical models suggest that NEC violations during inflation can amplify the primordial tensor power spectrum, leading to distinct features in the stochastic gravitational wave background (SGWB). In this work, we search for these NEC-violating signatures in the SGWB using data from Advanced LIGO and Advanced Virgo's first three observing runs. Our analysis reveals no statistically significant evidence of such signals, allowing us to place stringent upper limits on the tensor power spectrum amplitude, $P_{T,2}$, during the second inflationary stage. Specifically, we find that $P_{T,2} \lesssim 0.15$ at a $95\%$ confidence level. Notably, this upper limit is consistent with constraints derived from pulsar timing array observations, reinforcing the hypothesis that NEC violations during inflation could explain the signal detected by pulsar timing arrays. Our findings contribute to a deeper understanding of the early Universe and highlight the potential of current and future gravitational wave experiments in probing the physics of inflation and NEC violations.
}

\maketitle

\section{\label{sec:intro}Introduction}

The detection of gravitational waves (GWs) by Advanced LIGO~\cite{LIGOScientific:2014pky} and Advanced Virgo~\cite{VIRGO:2014yos} has marked the beginning of a new era in observational astronomy~\cite{LIGOScientific:2021djp}. The ability to detect GWs has opened up a new window for studying the Universe, providing invaluable insights into the physics of compact objects, such as black holes and neutron stars. In addition to individual, loud GW events from compact binary coalescences, the superposition of a large number of weaker, unresolved GW signals can form a stochastic gravitational-wave background (SGWB). SGWBs can carry valuable information about the properties and distribution of their sources, both astrophysical and cosmological.

Astrophysical sources of the SGWB include compact binary coalescences~\cite{Zhu:2011bd,Zhu:2012xw}, and core-collapse supernovae~\cite{Crocker:2017agi}. Cosmological sources, on the other hand, are related to various physical processes in the early Universe, such as cosmic phase transitions~\cite{Kibble:1980mv,Witten:1984rs,Mazumdar:2018dfl}, primordial density perturbations during inflation~\cite{Starobinsky:1979ty,Turner:1996ck,Bar-Kana:1994nri}, cosmic strings~\cite{Kibble:1976sj,Sarangi:2002yt,Damour:2004kw,Siemens:2006yp,LIGOScientific:2021nrg}, and cosmic domain walls~\cite{Vilenkin:1982ks,Sikivie:1982qv}. The detection of an SGWB of cosmological origin would provide unprecedented insights into the physics of the early Universe, potentially revealing new physics beyond the Standard Model.

Although the LIGO-Virgo Collaboration has not yet detected a stochastic gravitational-wave background (SGWB) signal during their first three observing runs, they have established upper limits on its amplitude~\cite{KAGRA:2021kbb}. These constraints are instrumental in testing the viability of various cosmological models that predict an amplified primordial GW spectrum. Notably, a class of such models postulates a violation of the null energy condition (NEC) during the inflationary epoch of the early Universe~\cite{Rubakov:2014jja,Cai:2016thi,Creminelli:2016zwa,Cai:2017tku}.

The NEC is a fundamental principle in general relativity, stating that for any null vector, the contraction with the energy-momentum tensor yields a non-negative value. Nevertheless, scenarios that involve NEC violation have been conjectured to account for a range of early Universe phenomena, including the inflationary epoch and the genesis of primordial black holes~\cite{Cai:2017dyi,Kolevatov:2017voe,Ilyas:2020zcb,Ilyas:2020qja,Zhu:2021ggm,Cai:2023uhc,Lesnefsky:2022fen,Easson:2024fzn}; see, for example, the review by~\cite{Rubakov:2014jja}. Advances in modified gravity theories, especially within the framework of ``beyond Horndeski'' models~\cite{Cai:2012va,Libanov:2016kfc,Kobayashi:2016xpl,Ijjas:2016vtq,Dobre:2017pnt,Zhu:2021whu,Cai:2022ori}, have shown that stable NEC violations are theoretically feasible, circumventing the issue of pathological instabilities.

The violation of the NEC has profound implications for the generation of primordial GWs in the early Universe. It has been shown that NEC violation can result in a notable amplification of the primordial tensor power spectrum at certain scales~\cite{Cai:2020qpu,Cai:2022nqv}. This amplification manifests as a blue tilt in the tensor power spectrum, indicating an enhanced production of high-frequency GWs. As a consequence, the SGWB is expected to demonstrate a distinctive pattern that reflects the NEC-violating phase of the early Universe.

The potential detection of this characteristic signature in the SGWB would provide compelling evidence for the existence of NEC-violating physics and shed light on the nature of gravity at high energies. Moreover, the study of NEC violation through its impact on the SGWB offers a novel approach to probing the physics of the early Universe, complementing other cosmological observations such as the cosmic microwave background (CMB) and large-scale structure surveys. By comparing theoretical predictions of NEC-violating models with observational constraints on the SGWB, we can gain valuable insights into the fundamental laws governing the Universe at its earliest stages and explore the possibilities of new physics beyond the standard cosmological paradigm.

In this paper, we investigate the constraints on NEC-violating inflationary models with GW data from Advanced LIGO and Advanced Virgo's first three observing runs. We focus on a specific scenario that incorporates an intermediate NEC-violating phase during inflation proposed in~\cite{Cai:2020qpu}. This model predicts a characteristic blue-tilted tensor power spectrum at scales corresponding to the NEC-violating phase, which can potentially be probed by current and future GW observations. The rest of this paper is structured as follows. In Section~\ref{NEC}, we begin by reviewing the key features of the inflationary model with an intermediate NEC-violating phase, focusing on its impact on the primordial tensor power spectrum. In Section~\ref{data}, we describe the methodology used to derive constraints on the model parameters using the data from Advanced LIGO and Advanced Virgo. Lastly, we summarise our results and give a discussion of the implications in Section~\ref{conclusion}.

\section{\label{NEC}SGWB from NEC violation during inflation}

In this section, we will provide a concise overview of the intermediate NEC violation during inflation. The scenario proposed by~\cite{Cai:2020qpu} describes the Universe starting with an initial phase of slow-roll inflation characterized by a Hubble parameter $H=H_\text{inf1}$. This is followed by a transition to a second phase of slow-roll inflation with a significantly larger Hubble parameter, $H=H_\text{inf2}$, after an intermediate stage of NEC violation. Throughout this evolution, the comoving Hubble horizon ($a^{-1}H^{-1}$) decreases, indicating the exit of perturbation modes from the horizon as the Universe undergoes accelerated expansion. These modes remain in the super-horizon regime until re-entering the horizon during the subsequent radiation-dominated or matter-dominated era.

Perturbation modes exiting the horizon during the first ($k < k_1$) and second ($k > k_2$) stages of slow-roll inflation have nearly scale-invariant power spectra, consistent with CMB temperature anisotropy observations. However, the power spectrum amplitude of modes exiting during the second stage is significantly larger. The scale-invariance of the tensor power spectrum at large scales ensures a non-suppressed tensor-to-scalar ratio $r$ and slow-roll parameter in canonical single-field slow-roll inflation. At small scales, the scale-invariance of both scalar and tensor power spectra prevents them from growing to $\mathcal{O}(1)$, preserving the validity of perturbation theory at higher frequencies. {In the specific scenario we have considered, the NEC violation occurs during an intermediate stage between two slow-roll inflationary phases. The end of inflation is attributed to the dynamics of the second slow-roll phase, which is assumed to follow the standard mechanism for ending inflation, such as the inflaton field reaching the minimum of its potential.}

The intermediate NEC violation during inflation can be realized with the effective field theory action~\cite{Cai:2016thi}:
\begin{equation}
S=\int d^4 x \sqrt{-g} \left[\frac{M_P^2}{2} R-\Lambda(t)-c(t) g^{00}
+ \frac{M_2^4(t)}{2}\left(\delta g^{00}\right)^2-\frac{m_3^3(t)}{2} \delta K \delta g^{00}+\frac{\tilde{m}_4^2(t)}{2} R^{(3)} \delta g^{00}\right].
\end{equation}
Here $\delta g^{00}=g^{00}+1$, $R^{(3)}$ denotes the Ricci scalar on the three-dimensional space-like hyper-surface, $\delta K=K-3H$, and $K$ is the extrinsic curvature.
The time-dependent functions $c(t)$ and $\Lambda(t)$ determine the background evolution through the relations $c(t) = -M_{\rm P}^{2} \dot{H}$ and $\Lambda(t) = M_{\rm P}^{2} (\dot{H} + 3 H^{2})$. The functions $M_2^4(t)$, $m_3^3(t)$, and $\tilde{m}_4^2(t)$ are determined or constrained to ensure that the scalar perturbations agree with observations.
{The EFT approach allows us to systematically incorporate modifications to general relativity and study their phenomenological consequences without committing to a specific underlying theory of quantum gravity. In the context of NEC violation, the EFT framework can accommodate such features by including higher-order operators that modify the standard kinetic and potential terms of the scalar field. These modifications can lead to scenarios where the usual energy conditions are not satisfied. Specifically, terms in the action such as $\frac{\tilde{m}_4^2(t)}{2} R^{(3)} \delta g^{00}$ and higher derivative interactions can effectively lead to negative contributions to the energy density, thus violating the NEC~\cite{Cai:2016thi}.}

Following~\cite{Ye:2023tpz}, we propose a parametrization of the primordial tensor power spectrum $P_T$ that captures the essential features:
\begin{equation}
P_{T} = P_{T,1} + \frac{\pi}{4}(2-n_T) \frac{k}{k_2}|g(k)|^2 P_{T,2} ~,\label{PT1211}
\end{equation}
where $P_{T,1}$ and $P_{T,2}$ represent the power spectrum amplitudes during the first and second slow-roll inflationary stages, respectively. The auxiliary function $g(k)$ is defined as
\begin{equation} \label{eq:g}
g(k) = H_{\frac{3-n_T}{2}}^{(1)} \left[ \frac{2-n_T}{2} \frac{k}{k_2} \right] \sin \frac{k}{k_2} + H_{\frac{1-n_T}{2}}^{(1)} \left[ \frac{2-n_T}{2} \frac{k}{k_2} \right]\left( \cos \frac{k}{k_2} - \frac{k_2}{k} \sin \frac{k}{k_2} \right) ~,
\end{equation}
where $H_{\nu}^{(1)}$ denotes the Hankel function of the first kind, and $n_T$ is the tensor spectral index during the NEC-violating stage. This parametrization has several key features:
\begin{itemize}
    \item For modes with wavenumbers $k\ll k_1$, which exit the horizon during the first slow-roll inflationary stage, the second term in Eq. (\ref{PT1211}) becomes negligible compared to $P_{T,1}$, and thus $P_T\approx P_{T,1}$. This ensures that the power spectrum at large scales is nearly scale-invariant and consistent with CMB observations.
    \item For modes with wavenumbers $k\gg k_1$, which exit the horizon during the second slow-roll inflationary stage or the NEC-violating stage, the second term in Eq. (\ref{PT1211}) dominates. This term captures the blue tilt and oscillatory features in the power spectrum around the transition scale $k\simeq k_2$, which are crucial for the generation of primordial black holes and the associated GW signals.
    \item Although the accuracy of this parametrization may be compromised around the first transition scale $k\simeq k_1$, the impact on our interests is minimal, as we focus on the features at intermediate and small scales.
\end{itemize}

A Gaussian, isotropic, unpolarized, and stationary SGWB is fully characterized by its spectral energy density, which is typically expressed in terms of the dimensionless quantity $\Omega_{\mathrm{GW}}(f)$. This quantity represents the GW energy density $\mathrm{d} \rho_{\text{GW}}$ within the frequency interval $f$ to $f+\mathrm{d} f$, multiplied by the GW frequency and divided by $\mathrm{d} f$ times the critical energy density $\rho_c$ required for a flat Universe:
\begin{equation}
\Omega_{\mathrm{GW}}(f)=\frac{f}{\rho_c} \frac{\mathrm{d} \rho_{\mathrm{GW}}}{\mathrm{d} f},
\end{equation}
where $\rho_c=3 H_0^2 c^2 /(8 \pi G)$, with $c$ being the speed of light and $G$ representing Newton's constant. The Hubble constant is taken from Planck 2018 observations as $H_0=67.4 \mathrm{~km} \mathrm{s}^{-1} \mathrm{Mpc}^{-1}$~\cite{Planck:2018vyg}.
To connect the primordial tensor power spectrum $P_T$ with the observed GW energy spectrum $\Omega_{\rm{GW}}$, we employ the following relation~\cite{Turner:1993vb}:
\begin{equation}
\begin{aligned}
\Omega_{\rm{GW}} &= \frac{k^2}{12 H_0^2}\left[\frac{3\Omega_mj_l(k\tau_0)}{k\tau_0}\sqrt{1.0 + 1.36\frac{k}{k_{\text{eq}}}+2.50\left(\frac{k}{k_{\text{eq}}}\right)^2}\right]^2P_T\\
&\simeq \frac{1}{24}\left(\frac{k}{H_0}\right)^2\left[\frac{3\Omega_m}{(k\tau_0)^2}\sqrt{1.0 + 1.36\frac{k}{k_{\text{eq}}}+2.50\left(\frac{k}{k_{\text{eq}}}\right)^2}\right]^2P_T
\end{aligned}
\end{equation}
where $\tau_{0}=1.41\times10^{4}$ Mpc is the conformal time today, $k_{\text{eq}}=0.073\,\Omega_{\text{m}} h^{2}$ Mpc$^{-1}$ is the wavenumber at matter-radiation equality, $\Omega_{\rm m}$ is the present matter density fraction, and the frequency $f$ is related to the wavenumber $k$ by $k=2\pi f$. This relation enables us to predict the observable GW energy spectrum based on our parametrization of the primordial tensor power spectrum, facilitating a direct comparison with current and future GW experiments.

\section{\label{data}Data analysis}
In this section, we describe the methodology employed to constrain the parameterized primordial tensor power spectrum, given by Eq.~\eqref{PT1211}, using the data from Advanced LIGO and Virgo's first three observing runs. As discussed in the previous section, we focus on the regime where $k_1\ll k$, in which the first term in Eq.~\eqref{PT1211} becomes negligible compared to the second term. This simplification allows us to safely neglect the contribution from $P_{T,1}$ and concentrate on the features arising from the NEC-violating stage and the second inflationary stage.
With this approximation, the theoretical spectrum given by Eq.~\eqref{PT1211} can be fully characterized by three key parameters: $n_T$, $f_c$, and $P_{T,2}$, where $f_c \equiv 2\pi k_2$.

To analyze the SGWB, we consider the LIGO-Hanford, LIGO-Livingston, and Virgo (HLV) network, with each detector labelled by the index $I={H, L, V}$. The time-series output of the detectors is denoted by $s_I(t)$, and its Fourier transform by $\tilde{s}_I(f)$. Following the formalism described in~\cite{Romano:2016dpx,Allen:1997ad}, we define the cross-correlation statistic for the baseline $IJ$ as
\begin{equation}\label{CIJ}
\hat{C}^{I J}(f)=\frac{2}{T} \frac{\operatorname{Re}[\tilde{s}I^{\star}(f) \tilde{s}J(f)]}{\gamma{I J}(f) S_0(f)},
\end{equation}
where $T$ represents the observation time, $\gamma{I J}(f)$ is the normalized overlap reduction function~\cite{Allen:1997ad} for the baseline $I J$, and the function $S_0(f)$ is defined as $S_0(f)=(3 H_0^2) /(10 \pi^2 f^3)$.
The estimator is normalized such that $\langle\hat{C}^{I J}(f)\rangle=\Omega_{\mathrm{GW}}(f)$ in the absence of correlated noise. In the limit of a small signal-to-noise ratio, the variance can be approximated as
\begin{equation}
\sigma_{I J}^2(f) \approx \frac{1}{2 T \Delta f} \frac{P_I(f) P_J(f)}{\gamma_{I J}^2(f) S_0^2(f)},
\end{equation}
where $\Delta f$ represents the frequency resolution, and $P_I(f)$ is the one-sided power spectral density in detector $I$.
\begin{table}[tbp]
\centering
\begin{tabular}{|c|c|c|}
\hline
Parameter& Description & Prior \\
\hline
$n_T$ & tensor spectral index during the NEC-violating stage & U$[0, 2]$ \\
$f_c\, (\mathrm{Hz})$ & transition frequency & LogU$[10^{-1}, 10^4]$ \\
$P_{T,2}$ &power spectrum amplitude during the second inflationary stage& LogU$[10^{-5}, 10^{2}]$ \\
\hline
\end{tabular}
\caption{\label{tab:prior}Prior distributions of the model parameters used in the Bayesian analysis, where U and LogU denote the uniform and log-uniform distributions, respectively.}
\end{table}
We perform a Bayesian analysis to search for the SGWB from the NEC-violating phase during inflation using data from the Advanced LIGO and Advanced Virgo's first three observing runs. The $\hat{C}^{I J}(f)$ data, which is model-independent and publicly available~\cite{KAGRA:2021kbb}, is utilized in this analysis. To estimate the parameters of a specific SGWB model, we combine the spectra from each baseline $I J$ to form the likelihood~\cite{Mandic:2012pj}
\begin{equation}\label{like}
p(\hat{C}^{I J}(f_k) | \boldsymbol{\theta}) \propto \exp \left[-\frac{1}{2} \sum_{I J} \sum_k\left(\frac{\hat{C}^{I J}(f_k)-\Omega_{\mathrm{M}}(f_k | \boldsymbol{\theta})}{\sigma_{I J}^2(f_k)}\right)^2\right],
\end{equation}
where we assume that the $\hat{C}^{I J}(f_k)$ follow a Gaussian distribution in the absence of a signal. The term $\Omega_{\mathrm{M}}(f | \boldsymbol{\theta})$ represents the SGWB model, characterized by the set of parameters $\boldsymbol{\theta}$ to be determined through the analysis. The likelihood is constructed by multiplying the contributions from all frequency bins and detector pairs.
Using the likelihood and Bayes' theorem, we form the posterior distribution $p(\boldsymbol{\theta} | C_k^{I J}) \propto p(C_k^{I J} | \boldsymbol{\theta}) p(\boldsymbol{\theta})$, where $p(\boldsymbol{\theta})$ represents the prior distribution on the parameters $\boldsymbol{\theta}$. To sample the parameter space and estimate the model parameters, we employ the Markov chain Monte Carlo method within the Bayesian framework.

\begin{figure}[tbp]
\centering
\includegraphics[width=0.8\textwidth]{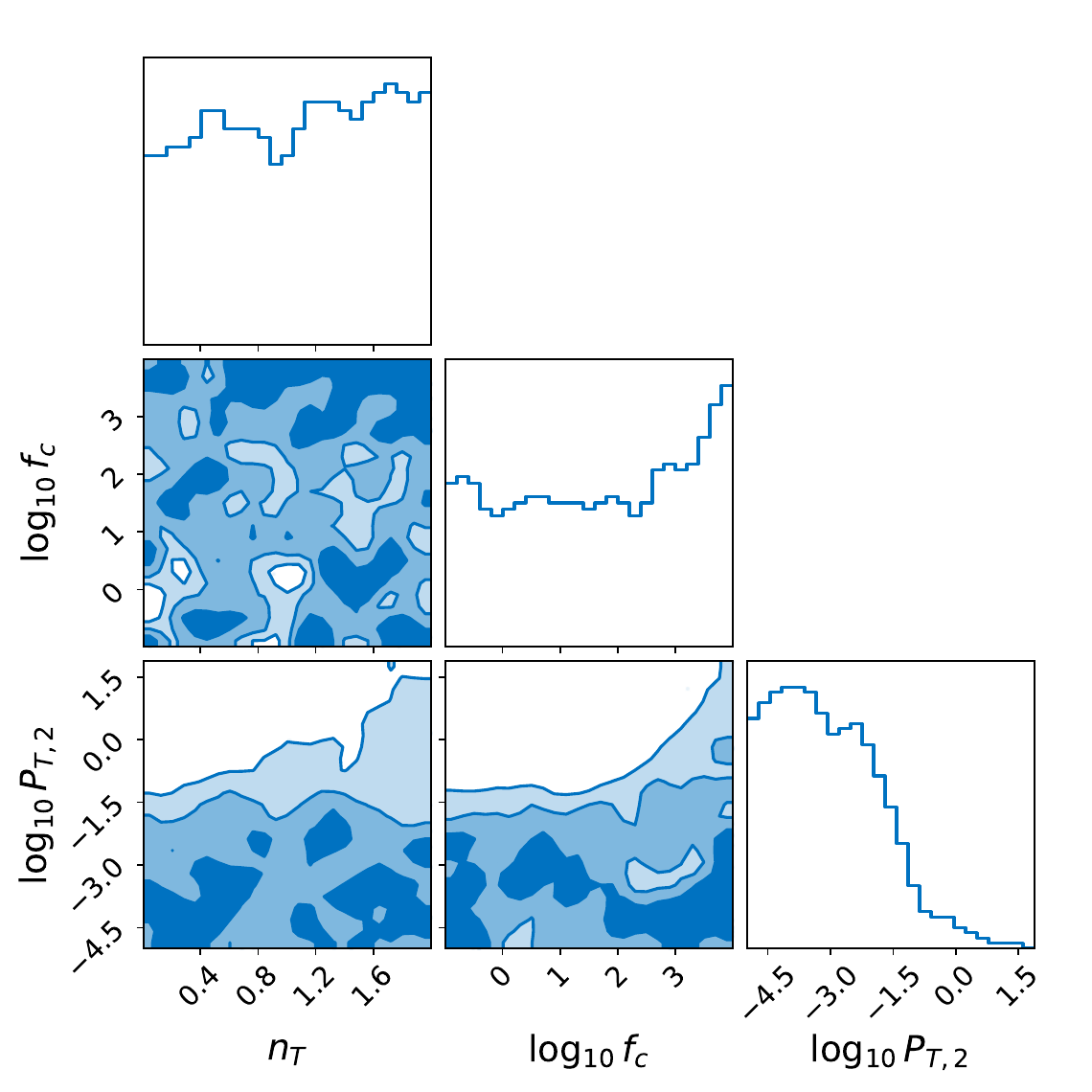}
\caption{\label{fig:posts}Posterior distributions for the model parameters $\boldsymbol{\theta}_{\mathrm{GW}} \equiv (n_T, f_c, P_{T,2})$. The contours represent the $1\sigma$, $2\sigma$, and $3\sigma$ confidence levels in the two-dimensional plots.}
\end{figure}

To quantify the statistical significance of the presence of the GW signal, we compute the Bayes factor between the model with SGWB signal and the model containing only noise, namely
\begin{equation}
\mathcal{B}_{\mathrm{NOISE}}^{\mathrm{GW}}=\frac{p(\hat{C}_{IJ} | \text{Model of signal})}{p(\hat{C}_{IJ} | \text{Pure noise})} = \frac{\int p(\hat{C}_{IJ}| \boldsymbol{\theta}_{\mathrm{GW}})\, p(\boldsymbol{\theta}_{\mathrm{GW}})\, \mathrm{d} \boldsymbol{\theta}_{\mathrm{GW}}}{\mathcal{N}}.
\end{equation}
Here, $\mathcal{N}$ is given by evaluating \Eq{like} for $\Omega_{\mathrm{M}}(f)=0$. When $\mathcal{B}_{\mathrm{NOISE}}^{\mathrm{GW}}>1$, there is support for the GW model compared to the NOISE model.
The free parameters in the analysis are $\boldsymbol{\theta}_{\mathrm{GW}} \equiv (n_T, f_c, P_{T,2})$, and their priors are listed in \Table{tab:prior}. The Bayesian analysis is performed using the \texttt{bilby} package~\cite{Ashton:2018jfp,Romero-Shaw:2020owr} and the dynamic nested sampling package \texttt{dynesty}~\cite{Speagle:2019ivv}.

\section{\label{conclusion}Results and discussions}
In this work, we have searched for the SGWB signal that could arise from a period of inflation in which the NEC was violated. Our analysis utilizes the combined cross-correlation spectrum of the first three observing runs from Advanced LIGO and Advanced Virgo, covering frequencies from $10\sim 200$ Hz. We present the posterior distributions of our model parameters in Figure~\ref{fig:posts}. Our findings demonstrate that the tensor spectral index $n_T$ and the transition frequency $f_c$, which are indicative of the NEC-violating phase, fall beyond the precise determination by the current instrumental capabilities of Advanced LIGO and Virgo.

To assess the presence of the SGWB signal originating from the NEC-violating phase, we calculate the Bayes factor $\mathcal{B}_{\text{NOISE}}^{\text{GW}}=0.53$. This value suggests that there is no significant evidence of such a signal in the current data. Consequently, we impose an upper limit on the power spectrum amplitude during the second inflationary stage, $P_{T,2}$. At a $95\%$ confidence level, we constrain $P_{T,2} \lesssim 0.15$. This bound may be applied to specific inflationary models that incorporate NEC violation, such as those featuring non-canonical kinetic terms~\cite{Creminelli:2006xe}, thereby restricting their viable parameter space.

Notably, our derived upper limit on $P_{T,2}$ is consistent with the constraints obtained from PTA observations~\cite{Ye:2023tpz}. This consistency highlights the complementary nature of PTA and Advanced LIGO-Virgo observations in probing the physics of the early Universe. PTAs are sensitive to GWs in the nanohertz frequency range, while Advanced LIGO-Virgo explores higher frequencies between 10 and 200 Hz. The agreement between the constraints obtained from these different frequency regimes strengthens the case for the NEC-violating inflationary scenario as a plausible explanation for the PTA signal. As future GW experiments with improved sensitivities come online, we anticipate enhanced capabilities in testing inflationary models and further constraining their parameters.

In summary, our analysis demonstrates that the current Advanced LIGO and Advanced Virgo data can effectively constrain the primordial tensor power spectrum in the presence of an NEC-violating phase during inflation. The obtained upper limit on $P_{T,2}$ sheds light on the properties of the NEC-violating stage and the subsequent transition to the second inflationary phase. These findings contribute to our understanding of the early Universe and emphasize the potential of future GW experiments in probing the physics of inflation and NEC violation. 

\acknowledgments
ZCC is supported by the National Natural Science Foundation of China (Grant No.~12247176 and No.~12247112) and the innovative research group of Hunan Province under Grant No. 2024JJ1006.
LL is supported by the National Natural Science Foundation of China (Grant No. 12247112 and No. 12247176) and the China Postdoctoral Science Foundation Fellowship No. 2023M730300.

\bibliographystyle{JHEP}
\bibliography{ref}

\providecommand{\href}[2]{#2}\begingroup\raggedright\begin{thebibliography}{10}

\bibitem{LIGOScientific:2014pky}
{\scshape LIGO Scientific} collaboration, \emph{{Advanced LIGO}}, \href{https://doi.org/10.1088/0264-9381/32/7/074001}{\emph{Class. Quant. Grav.} {\bfseries 32} (2015) 074001} [\href{https://arxiv.org/abs/1411.4547}{{\ttfamily 1411.4547}}].

\bibitem{VIRGO:2014yos}
{\scshape VIRGO} collaboration, \emph{{Advanced Virgo: a second-generation interferometric gravitational wave detector}}, \href{https://doi.org/10.1088/0264-9381/32/2/024001}{\emph{Class. Quant. Grav.} {\bfseries 32} (2015) 024001} [\href{https://arxiv.org/abs/1408.3978}{{\ttfamily 1408.3978}}].

\bibitem{LIGOScientific:2021djp}
{\scshape KAGRA, VIRGO, LIGO Scientific} collaboration, \emph{{GWTC-3: Compact Binary Coalescences Observed by LIGO and Virgo during the Second Part of the Third Observing Run}}, \href{https://doi.org/10.1103/PhysRevX.13.041039}{\emph{Phys. Rev. X} {\bfseries 13} (2023) 041039} [\href{https://arxiv.org/abs/2111.03606}{{\ttfamily 2111.03606}}].

\bibitem{Zhu:2011bd}
X.-J.~Zhu, E.~Howell, T.~Regimbau, D.~Blair and Z.-H.~Zhu, \emph{{Stochastic Gravitational Wave Background from Coalescing Binary Black Holes}}, \href{https://doi.org/10.1088/0004-637X/739/2/86}{\emph{Astrophys. J.} {\bfseries 739} (2011) 86} [\href{https://arxiv.org/abs/1104.3565}{{\ttfamily 1104.3565}}].

\bibitem{Zhu:2012xw}
X.-J.~Zhu, E.J.~Howell, D.G.~Blair and Z.-H.~Zhu, \emph{{On the gravitational wave background from compact binary coalescences in the band of ground-based interferometers}}, \href{https://doi.org/10.1093/mnras/stt207}{\emph{Mon. Not. Roy. Astron. Soc.} {\bfseries 431} (2013) 882} [\href{https://arxiv.org/abs/1209.0595}{{\ttfamily 1209.0595}}].

\bibitem{Crocker:2017agi}
K.~Crocker, T.~Prestegard, V.~Mandic, T.~Regimbau, K.~Olive and E.~Vangioni, \emph{{Systematic study of the stochastic gravitational-wave background due to stellar core collapse}}, \href{https://doi.org/10.1103/PhysRevD.95.063015}{\emph{Phys. Rev. D} {\bfseries 95} (2017) 063015} [\href{https://arxiv.org/abs/1701.02638}{{\ttfamily 1701.02638}}].

\bibitem{Kibble:1980mv}
T.W.B.~Kibble, \emph{{Some Implications of a Cosmological Phase Transition}}, \href{https://doi.org/10.1016/0370-1573(80)90091-5}{\emph{Phys. Rept.} {\bfseries 67} (1980) 183}.

\bibitem{Witten:1984rs}
E.~Witten, \emph{{Cosmic Separation of Phases}}, \href{https://doi.org/10.1103/PhysRevD.30.272}{\emph{Phys. Rev. D} {\bfseries 30} (1984) 272}.

\bibitem{Mazumdar:2018dfl}
A.~Mazumdar and G.~White, \emph{{Review of cosmic phase transitions: their significance and experimental signatures}}, \href{https://doi.org/10.1088/1361-6633/ab1f55}{\emph{Rept. Prog. Phys.} {\bfseries 82} (2019) 076901} [\href{https://arxiv.org/abs/1811.01948}{{\ttfamily 1811.01948}}].

\bibitem{Starobinsky:1979ty}
A.A.~Starobinsky, \emph{{Spectrum of relict gravitational radiation and the early state of the universe}}, {\emph{JETP Lett.} {\bfseries 30} (1979) 682}.

\bibitem{Turner:1996ck}
M.S.~Turner, \emph{{Detectability of inflation produced gravitational waves}}, \href{https://doi.org/10.1103/PhysRevD.55.R435}{\emph{Phys. Rev. D} {\bfseries 55} (1997) R435} [\href{https://arxiv.org/abs/astro-ph/9607066}{{\ttfamily astro-ph/9607066}}].

\bibitem{Bar-Kana:1994nri}
R.~Bar-Kana, \emph{{Limits on direct detection of gravitational waves}}, \href{https://doi.org/10.1103/PhysRevD.50.1157}{\emph{Phys. Rev. D} {\bfseries 50} (1994) 1157} [\href{https://arxiv.org/abs/astro-ph/9401050}{{\ttfamily astro-ph/9401050}}].

\bibitem{Kibble:1976sj}
T.W.B.~Kibble, \emph{{Topology of Cosmic Domains and Strings}}, \href{https://doi.org/10.1088/0305-4470/9/8/029}{\emph{J. Phys. A} {\bfseries 9} (1976) 1387}.

\bibitem{Sarangi:2002yt}
S.~Sarangi and S.H.H.~Tye, \emph{{Cosmic string production towards the end of brane inflation}}, \href{https://doi.org/10.1016/S0370-2693(02)01824-5}{\emph{Phys. Lett. B} {\bfseries 536} (2002) 185} [\href{https://arxiv.org/abs/hep-th/0204074}{{\ttfamily hep-th/0204074}}].

\bibitem{Damour:2004kw}
T.~Damour and A.~Vilenkin, \emph{{Gravitational radiation from cosmic (super)strings: Bursts, stochastic background, and observational windows}}, \href{https://doi.org/10.1103/PhysRevD.71.063510}{\emph{Phys. Rev. D} {\bfseries 71} (2005) 063510} [\href{https://arxiv.org/abs/hep-th/0410222}{{\ttfamily hep-th/0410222}}].

\bibitem{Siemens:2006yp}
X.~Siemens, V.~Mandic and J.~Creighton, \emph{{Gravitational wave stochastic background from cosmic (super)strings}}, \href{https://doi.org/10.1103/PhysRevLett.98.111101}{\emph{Phys. Rev. Lett.} {\bfseries 98} (2007) 111101} [\href{https://arxiv.org/abs/astro-ph/0610920}{{\ttfamily astro-ph/0610920}}].

\bibitem{LIGOScientific:2021nrg}
{\scshape LIGO Scientific, Virgo, KAGRA} collaboration, \emph{{Constraints on Cosmic Strings Using Data from the Third Advanced LIGO\textendash{}Virgo Observing Run}}, \href{https://doi.org/10.1103/PhysRevLett.126.241102}{\emph{Phys. Rev. Lett.} {\bfseries 126} (2021) 241102} [\href{https://arxiv.org/abs/2101.12248}{{\ttfamily 2101.12248}}].

\bibitem{Vilenkin:1982ks}
A.~Vilenkin and A.E.~Everett, \emph{{Cosmic Strings and Domain Walls in Models with Goldstone and PseudoGoldstone Bosons}}, \href{https://doi.org/10.1103/PhysRevLett.48.1867}{\emph{Phys. Rev. Lett.} {\bfseries 48} (1982) 1867}.

\bibitem{Sikivie:1982qv}
P.~Sikivie, \emph{{Of Axions, Domain Walls and the Early Universe}}, \href{https://doi.org/10.1103/PhysRevLett.48.1156}{\emph{Phys. Rev. Lett.} {\bfseries 48} (1982) 1156}.

\bibitem{KAGRA:2021kbb}
{\scshape KAGRA, Virgo, LIGO Scientific} collaboration, \emph{{Upper limits on the isotropic gravitational-wave background from Advanced LIGO and Advanced Virgo\textquoteright{}s third observing run}}, \href{https://doi.org/10.1103/PhysRevD.104.022004}{\emph{Phys. Rev. D} {\bfseries 104} (2021) 022004} [\href{https://arxiv.org/abs/2101.12130}{{\ttfamily 2101.12130}}].

\bibitem{Rubakov:2014jja}
V.A.~Rubakov, \emph{{The Null Energy Condition and its violation}}, \href{https://doi.org/10.3367/UFNe.0184.201402b.0137}{\emph{Phys. Usp.} {\bfseries 57} (2014) 128} [\href{https://arxiv.org/abs/1401.4024}{{\ttfamily 1401.4024}}].

\bibitem{Cai:2016thi}
Y.~Cai, Y.~Wan, H.-G.~Li, T.~Qiu and Y.-S.~Piao, \emph{{The Effective Field Theory of nonsingular cosmology}}, \href{https://doi.org/10.1007/JHEP01(2017)090}{\emph{JHEP} {\bfseries 01} (2017) 090} [\href{https://arxiv.org/abs/1610.03400}{{\ttfamily 1610.03400}}].

\bibitem{Creminelli:2016zwa}
P.~Creminelli, D.~Pirtskhalava, L.~Santoni and E.~Trincherini, \emph{{Stability of Geodesically Complete Cosmologies}}, \href{https://doi.org/10.1088/1475-7516/2016/11/047}{\emph{JCAP} {\bfseries 11} (2016) 047} [\href{https://arxiv.org/abs/1610.04207}{{\ttfamily 1610.04207}}].

\bibitem{Cai:2017tku}
Y.~Cai, H.-G.~Li, T.~Qiu and Y.-S.~Piao, \emph{{The Effective Field Theory of nonsingular cosmology: II}}, \href{https://doi.org/10.1140/epjc/s10052-017-4938-y}{\emph{Eur. Phys. J. C} {\bfseries 77} (2017) 369} [\href{https://arxiv.org/abs/1701.04330}{{\ttfamily 1701.04330}}].

\bibitem{Cai:2017dyi}
Y.~Cai and Y.-S.~Piao, \emph{{A covariant Lagrangian for stable nonsingular bounce}}, \href{https://doi.org/10.1007/JHEP09(2017)027}{\emph{JHEP} {\bfseries 09} (2017) 027} [\href{https://arxiv.org/abs/1705.03401}{{\ttfamily 1705.03401}}].

\bibitem{Kolevatov:2017voe}
R.~Kolevatov, S.~Mironov, N.~Sukhov and V.~Volkova, \emph{{Cosmological bounce and Genesis beyond Horndeski}}, \href{https://doi.org/10.1088/1475-7516/2017/08/038}{\emph{JCAP} {\bfseries 08} (2017) 038} [\href{https://arxiv.org/abs/1705.06626}{{\ttfamily 1705.06626}}].

\bibitem{Ilyas:2020zcb}
A.~Ilyas, M.~Zhu, Y.~Zheng and Y.-F.~Cai, \emph{{Emergent Universe and Genesis from the DHOST Cosmology}}, \href{https://doi.org/10.1007/JHEP01(2021)141}{\emph{JHEP} {\bfseries 01} (2021) 141} [\href{https://arxiv.org/abs/2009.10351}{{\ttfamily 2009.10351}}].

\bibitem{Ilyas:2020qja}
A.~Ilyas, M.~Zhu, Y.~Zheng, Y.-F.~Cai and E.N.~Saridakis, \emph{{DHOST Bounce}}, \href{https://doi.org/10.1088/1475-7516/2020/09/002}{\emph{JCAP} {\bfseries 09} (2020) 002} [\href{https://arxiv.org/abs/2002.08269}{{\ttfamily 2002.08269}}].

\bibitem{Zhu:2021ggm}
M.~Zhu and Y.~Zheng, \emph{{Improved DHOST Genesis}}, \href{https://doi.org/10.1007/JHEP11(2021)163}{\emph{JHEP} {\bfseries 11} (2021) 163} [\href{https://arxiv.org/abs/2109.05277}{{\ttfamily 2109.05277}}].

\bibitem{Cai:2023uhc}
Y.~Cai, M.~Zhu and Y.-S.~Piao, \emph{{Primordial black holes from null energy condition violation during inflation}},  \href{https://arxiv.org/abs/2305.10933}{{\ttfamily 2305.10933}}.

\bibitem{Lesnefsky:2022fen}
J.E.~Lesnefsky, D.A.~Easson and P.C.W.~Davies, \emph{{Past-completeness of inflationary spacetimes}}, \href{https://doi.org/10.1103/PhysRevD.107.044024}{\emph{Phys. Rev. D} {\bfseries 107} (2023) 044024} [\href{https://arxiv.org/abs/2207.00955}{{\ttfamily 2207.00955}}].

\bibitem{Easson:2024fzn}
D.A.~Easson and J.E.~Lesnefsky, \emph{{Eternal Universes}},  \href{https://arxiv.org/abs/2404.03016}{{\ttfamily 2404.03016}}.

\bibitem{Cai:2012va}
Y.-F.~Cai, D.A.~Easson and R.~Brandenberger, \emph{{Towards a Nonsingular Bouncing Cosmology}}, \href{https://doi.org/10.1088/1475-7516/2012/08/020}{\emph{JCAP} {\bfseries 08} (2012) 020} [\href{https://arxiv.org/abs/1206.2382}{{\ttfamily 1206.2382}}].

\bibitem{Libanov:2016kfc}
M.~Libanov, S.~Mironov and V.~Rubakov, \emph{{Generalized Galileons: instabilities of bouncing and Genesis cosmologies and modified Genesis}}, \href{https://doi.org/10.1088/1475-7516/2016/08/037}{\emph{JCAP} {\bfseries 08} (2016) 037} [\href{https://arxiv.org/abs/1605.05992}{{\ttfamily 1605.05992}}].

\bibitem{Kobayashi:2016xpl}
T.~Kobayashi, \emph{{Generic instabilities of nonsingular cosmologies in Horndeski theory: A no-go theorem}}, \href{https://doi.org/10.1103/PhysRevD.94.043511}{\emph{Phys. Rev. D} {\bfseries 94} (2016) 043511} [\href{https://arxiv.org/abs/1606.05831}{{\ttfamily 1606.05831}}].

\bibitem{Ijjas:2016vtq}
A.~Ijjas and P.J.~Steinhardt, \emph{{Fully stable cosmological solutions with a non-singular classical bounce}}, \href{https://doi.org/10.1016/j.physletb.2016.11.047}{\emph{Phys. Lett. B} {\bfseries 764} (2017) 289} [\href{https://arxiv.org/abs/1609.01253}{{\ttfamily 1609.01253}}].

\bibitem{Dobre:2017pnt}
D.A.~Dobre, A.V.~Frolov, J.T.~G\'alvez~Ghersi, S.~Ramazanov and A.~Vikman, \emph{{Unbraiding the Bounce: Superluminality around the Corner}}, \href{https://doi.org/10.1088/1475-7516/2018/03/020}{\emph{JCAP} {\bfseries 03} (2018) 020} [\href{https://arxiv.org/abs/1712.10272}{{\ttfamily 1712.10272}}].

\bibitem{Zhu:2021whu}
M.~Zhu, A.~Ilyas, Y.~Zheng, Y.-F.~Cai and E.N.~Saridakis, \emph{{Scalar and tensor perturbations in DHOST bounce cosmology}}, \href{https://doi.org/10.1088/1475-7516/2021/11/045}{\emph{JCAP} {\bfseries 11} (2021) 045} [\href{https://arxiv.org/abs/2108.01339}{{\ttfamily 2108.01339}}].

\bibitem{Cai:2022ori}
Y.~Cai, J.~Xu, S.~Zhao and S.~Zhou, \emph{{Perturbative unitarity and NEC violation in genesis cosmology}}, \href{https://doi.org/10.1007/JHEP10(2022)140}{\emph{JHEP} {\bfseries 10} (2022) 140} [\href{https://arxiv.org/abs/2207.11772}{{\ttfamily 2207.11772}}].

\bibitem{Cai:2020qpu}
Y.~Cai and Y.-S.~Piao, \emph{{Intermittent null energy condition violations during inflation and primordial gravitational waves}}, \href{https://doi.org/10.1103/PhysRevD.103.083521}{\emph{Phys. Rev. D} {\bfseries 103} (2021) 083521} [\href{https://arxiv.org/abs/2012.11304}{{\ttfamily 2012.11304}}].

\bibitem{Cai:2022nqv}
Y.~Cai and Y.-S.~Piao, \emph{{Generating enhanced primordial GWs during inflation with intermittent violation of NEC and diminishment of GW propagating speed}}, \href{https://doi.org/10.1007/JHEP06(2022)067}{\emph{JHEP} {\bfseries 06} (2022) 067} [\href{https://arxiv.org/abs/2201.04552}{{\ttfamily 2201.04552}}].

\bibitem{Ye:2023tpz}
G.~Ye, M.~Zhu and Y.~Cai, \emph{{Null energy condition violation during inflation and pulsar timing array observations}}, \href{https://doi.org/10.1007/JHEP02(2024)008}{\emph{JHEP} {\bfseries 02} (2024) 008} [\href{https://arxiv.org/abs/2312.10685}{{\ttfamily 2312.10685}}].

\bibitem{Planck:2018vyg}
{\scshape Planck} collaboration, \emph{{Planck 2018 results. VI. Cosmological parameters}}, \href{https://doi.org/10.1051/0004-6361/201833910}{\emph{Astron. Astrophys.} {\bfseries 641} (2020) A6} [\href{https://arxiv.org/abs/1807.06209}{{\ttfamily 1807.06209}}].

\bibitem{Turner:1993vb}
M.S.~Turner, M.J.~White and J.E.~Lidsey, \emph{{Tensor perturbations in inflationary models as a probe of cosmology}}, \href{https://doi.org/10.1103/PhysRevD.48.4613}{\emph{Phys. Rev. D} {\bfseries 48} (1993) 4613} [\href{https://arxiv.org/abs/astro-ph/9306029}{{\ttfamily astro-ph/9306029}}].

\bibitem{Romano:2016dpx}
J.D.~Romano and N.J.~Cornish, \emph{{Detection methods for stochastic gravitational-wave backgrounds: a unified treatment}}, \href{https://doi.org/10.1007/s41114-017-0004-1}{\emph{Living Rev. Rel.} {\bfseries 20} (2017) 2} [\href{https://arxiv.org/abs/1608.06889}{{\ttfamily 1608.06889}}].

\bibitem{Allen:1997ad}
B.~Allen and J.D.~Romano, \emph{{Detecting a stochastic background of gravitational radiation: Signal processing strategies and sensitivities}}, \href{https://doi.org/10.1103/PhysRevD.59.102001}{\emph{Phys. Rev. D} {\bfseries 59} (1999) 102001} [\href{https://arxiv.org/abs/gr-qc/9710117}{{\ttfamily gr-qc/9710117}}].

\bibitem{Mandic:2012pj}
V.~Mandic, E.~Thrane, S.~Giampanis and T.~Regimbau, \emph{{Parameter Estimation in Searches for the Stochastic Gravitational-Wave Background}}, \href{https://doi.org/10.1103/PhysRevLett.109.171102}{\emph{Phys. Rev. Lett.} {\bfseries 109} (2012) 171102} [\href{https://arxiv.org/abs/1209.3847}{{\ttfamily 1209.3847}}].

\bibitem{Ashton:2018jfp}
G.~Ashton et~al., \emph{{BILBY: A user-friendly Bayesian inference library for gravitational-wave astronomy}}, \href{https://doi.org/10.3847/1538-4365/ab06fc}{\emph{Astrophys. J. Suppl.} {\bfseries 241} (2019) 27} [\href{https://arxiv.org/abs/1811.02042}{{\ttfamily 1811.02042}}].

\bibitem{Romero-Shaw:2020owr}
I.M.~Romero-Shaw et~al., \emph{{Bayesian inference for compact binary coalescences with bilby: validation and application to the first LIGO\textendash{}Virgo gravitational-wave transient catalogue}}, \href{https://doi.org/10.1093/mnras/staa2850}{\emph{Mon. Not. Roy. Astron. Soc.} {\bfseries 499} (2020) 3295} [\href{https://arxiv.org/abs/2006.00714}{{\ttfamily 2006.00714}}].

\bibitem{Speagle:2019ivv}
J.S.~Speagle, \emph{{dynesty: a dynamic nested sampling package for estimating Bayesian posteriors and evidences}}, \href{https://doi.org/10.1093/mnras/staa278}{\emph{Mon. Not. Roy. Astron. Soc.} {\bfseries 493} (2020) 3132} [\href{https://arxiv.org/abs/1904.02180}{{\ttfamily 1904.02180}}].

\bibitem{Creminelli:2006xe}
P.~Creminelli, M.A.~Luty, A.~Nicolis and L.~Senatore, \emph{{Starting the Universe: Stable Violation of the Null Energy Condition and Non-standard Cosmologies}}, \href{https://doi.org/10.1088/1126-6708/2006/12/080}{\emph{JHEP} {\bfseries 12} (2006) 080} [\href{https://arxiv.org/abs/hep-th/0606090}{{\ttfamily hep-th/0606090}}].

\end{thebibliography}\endgroup

\end{document}